\pdfoutput=1
\documentclass{iopart}

\usepackage{iopams}
\usepackage{subfig}
\usepackage{graphicx}
\usepackage{color}

\expandafter\let\csname equation*\endcsname\relax
\expandafter\let\csname endequation*\endcsname\relax
\usepackage{amsmath}

\newcommand{\beq}{\begin{equation}}
\newcommand{\eeq}{\end{equation}}
\newcommand{\bdm}{\begin{displaymath}}
\newcommand{\edm}{\end{displaymath}}

\begin{document}

\title[]
{Optimization of seismometer arrays for the cancellation of Newtonian noise from seismic body waves}

\author{F Badaracco and J Harms}
\vskip 1mm
\address{Gran Sasso Science Institute (GSSI), I-67100 L'Aquila, Italy}
\address{INFN, Laboratori Nazionali del Gran Sasso, I-67100 Assergi, Italy}

\begin{abstract}
Newtonian noise from seismic fields is predicted to become a sensitivity limiting noise contribution of the gravitational-wave detectors Advanced LIGO and Virgo in the next few years. It also plays a major role in the planning of next-generation detectors, which might be constructed underground, as planned for the Einstein Telescope, mostly to suppress Newtonian noise. Coherent noise cancellation using Wiener filters provides a way to mitigate Newtonian noise. So far, only the cancellation of Newtonian noise produced by seismic surface waves has been studied in detail due to its relevance for Advanced LIGO and Virgo. However, seismic body waves can still contribute significantly to Newtonian noise in surface detectors, and they might be the dominant source of gravity fluctuations in underground detectors. In this paper, we present the first detailed analysis of coherent cancellation of Newtonian noise from body waves. While the required number of seismometers to achieve a certain level of noise suppression is higher than for seismic surface waves, we show that optimal seismometer arrays can greatly reduce body-wave Newtonian noise. The optimal array configurations and achieved residuals depend strongly on the composition of the seismic field in terms of average compressional-wave and shear-wave content. We propose Newtonian-noise cancellation to achieve the ambitious low-frequency target of the Einstein Telescope. 
\end{abstract}


\section{Introduction}
Terrestrial gravity fluctuations contribute to the instrumental noise in ground-based gravitational-wave (GW) detectors as so-called Newtonian noise (NN) \cite{Sau1984}. It is one of the noise sources originally considered a facility limit of GW detectors since it was assumed that only major facility modifications could mitigate this noise. In the meantime, the conception of NN has changed. Coherent noise cancellation using Wiener filters is being developed as a means to suppress NN \cite{Cel2000,CoEA2016a}. The idea is to use environmental monitors such as seismometers and microphones to produce a coherent estimate of the associated gravity fluctuations. Newtonian noise produced by seismic fields is expected to become a sensitivity limit in the Advanced LIGO \cite{LSC2015} and Virgo \cite{AcEA2015} detectors between about 10\,Hz and 30\,Hz \cite{DHA2012,Har2015}. In addition, contributions from an atmospheric sound field can be significant \cite{FiEA2018}. 

Seismic fields are composed of body and surface waves \cite{AkRi2009}. Analyses at the LIGO Hanford detector showed that surface displacement is dominated by Rayleigh waves \cite{CoEA2018a} (similar unpublished results were obtained for the Virgo detector). This result was anticipated since the dominant ground vibrations are produced by local seismic sources located at the surface such as ventilation fans and pumps. Consequently, the development of NN cancellation systems for LIGO and Virgo focuses on NN from Rayleigh waves \cite{CoEA2016a}. Contributions from body waves are being neglected. 

The situation changes once a detector upgrade relies on high suppression of NN by a factor 10 or more. In this case, relatively weak body-wave NN might become significant. More importantly even, body-wave NN might be the dominant contribution to NN in underground GW detectors such as KAGRA \cite{AkEA2018} or the planned Einstein Telescope \cite{PuEA2010}, where the main incentive to build a GW detector underground is to strongly suppress NN from atmospheric and seismic surface fields \cite{BeEA2010,FiEA2018}. Since first NN estimates for the Einstein Telescope neglected contributions from body waves, it was not immediately realized that NN cancellation will still be required to reach the low-frequency sensitivity target as shown for example in \cite{HiEA2011}. Educated guessing of underground array configurations to achieve body-wave NN cancellation did not lead to satisfactory results \cite{Har2015}. It is therefore necessary to search for optimal array configurations and understand how these depend on properties of the seismic field.

In this paper, we investigate the performance of optimized seismometer arrays for the cancellation of NN from body waves. We consider a test mass sufficiently far underground so that the seismometers can be placed anywhere around the test mass up to distances of a few 100\,m. Body waves can be shear or compressional waves. Both produce NN through displacement of a cavity wall, which hosts the test mass of the GW detector. Compressional waves produce additional NN through compression and dilation of rock. In section \ref{sec:NNcanc}, we present the correlation functions of an isotropic seismic field required to calculate the Wiener filter. In section \ref{sec:optim}, we present our solutions of optimized arrays. Cancellation performance is investigated as a function of the number of seismometers, and on the compressional-wave to shear-wave content  ratio. Furthermore, it is studied how sensitive the performance is to the exact placement of seismometers. Implications for the Einstein Telescope are discussed in section \ref{sec:ET}. We then conclude in section \ref{sec:concl}.

\section{Wiener filters for underground NN cancellation}
\label{sec:NNcanc}
In underground detectors, test masses will be located in cavities fully surrounded by hard rock. Seismic waves propagating through the rock will cause NN. A possible mitigation strategy is to cancel part of the NN using an optimal linear filter, the Wiener filter \cite{Vas2001,Homestake}, which provides a coherent estimate of NN from seismic observations. We assume that NN picked up by different test masses is uncorrelated, which means that the detector arms need to be long (several kilometers depending on seismic speeds), and the arms need to be perpendicular and the seismic field isotropic (otherwise, NN between the two inner test masses located near the vertex of an interferometer can be correlated). In this case, one can cancel all of a detector's NN by canceling NN from each test mass individually. It should be noted though that the seismic measurements for NN cancellation at the two inner test masses will in any case be correlated, which means that the Wiener filter will be different, but it will perform equally well or better than predicted by the single test-mass analysis.

Following Newton's law, we can write the perturbation of gravity acceleration as follows:
\begin{equation}
\delta \vec{a}\left(\vec{r}_0,t\right) = -G \int dV \rho\left(\vec{r}\right)\left(\vec{\xi}\left(\vec{r},t\right)\cdot \nabla_0\right)\frac{\vec{r}-\vec{r}_0}{\left| \vec{r} -\vec{r}_0\right|^3}
\label{eq:NNbase}
\end{equation}
where $\vec{\xi}(\vec{r},t)$ is the seismic displacement field. The linear dependence of the gravity perturbation on the displacement field makes it explicit that correlations between seismic displacement and NN must exist. These correlations determine the Wiener filter. Wiener filters can be formulated in time and frequency domain. For Gaussian, stationary noise as considered throughout this paper, frequency-domain correlations are expressed as cross-spectral densities (CSDs) \cite{Har2015}. The performance of a Wiener filter can be quantified by the relative residual NN spectral density spectrum $R(\omega)$ that it leaves in the GW data \cite{Cel2000}:
\begin{equation}
R(\omega) = 1 - \frac{\vec{C}^\dagger_{\rm SN}\left(\omega\right)\cdot\left(\vec{C}_{\rm SS}(\omega)\right)^{-1}\cdot \vec{C}_{\rm SN}(\omega)}{C_{\rm NN}(\omega)}
\label{eq:residual}
\end{equation}
Here, $\vec{C}_{\rm SN}$ represents the vector of CSDs between the displacement recorded by $N$ seismometers and NN, $C_{\rm SS}$ is the matrix of CSDs between all seismometers, and $C_{\rm NN}$ is NN spectral density. The best possible cancellation using $N$ equal seismometers characterized by a certain signal-to-noise ratio (SNR) is achieved if the seismometers' data are all exact copies (up to some irrelevant transfer function) of the NN so that the CSD between NN and seismometers assumes its theoretical maximum. In this case, the noise residual is
\begin{equation}
R_{\rm min}(\omega) = 1-\frac{1}{1+1/(N\cdot{\rm SNR}(\omega)^2)}\approx\frac{1}{N\cdot{\rm SNR}(\omega)^2}.
\label{eq:minR}
\end{equation}
For sufficiently high $N$ so that the residual $R$ is limited by the SNR of the sensors, the residual $R$ from optimal arrays needs to fall at least with $1/N$, since one can always just add a new seismometer next to an existing one effectively averaging over seismometer instrument noise. 

We will only consider the cancellation of NN from a single test mass. The residual $R$ can be understood as a function of the seismometer positions with a fixed number of seismometers. One can then search for the seismometer positions that minimize the residual. We choose here to optimize the array for a fixed frequency, which translates into a fixed length of the seismic waves. 

\begin{figure}
\centering{
\subfloat[p = 1]{\includegraphics[width=0.4\textwidth]{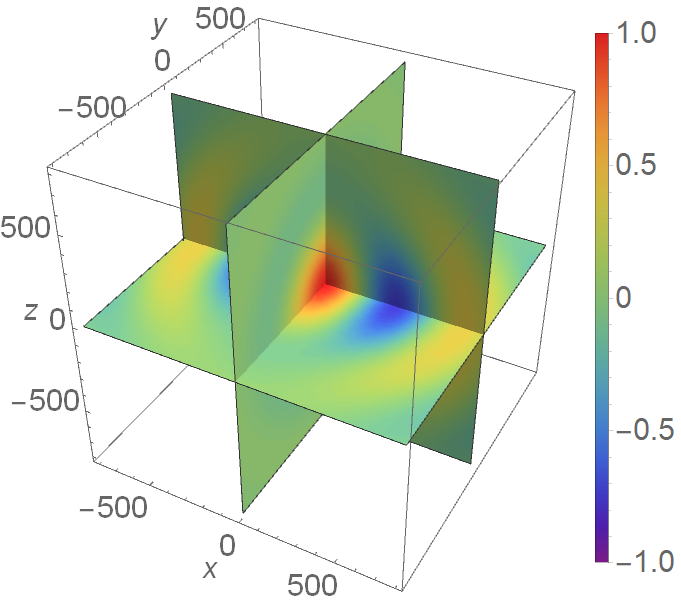}}
\subfloat[p = 1/3]{\includegraphics[width=0.4\textwidth]{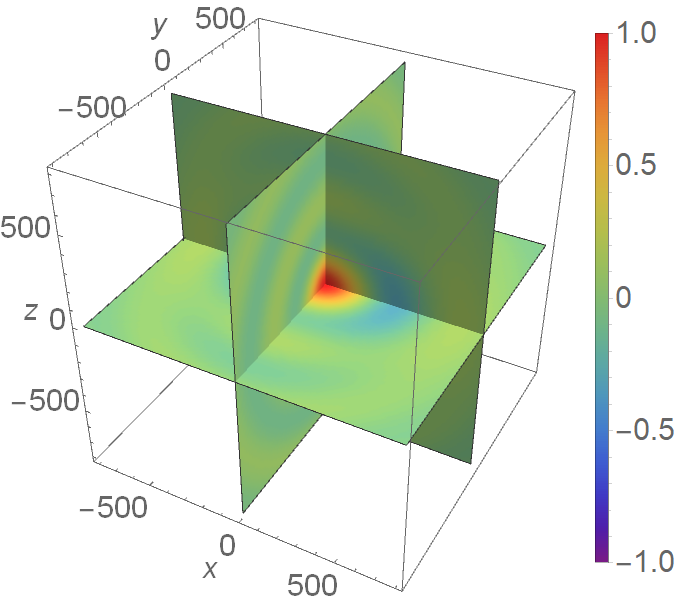}}}
\caption{Normalized CSDs between two seismometers in isotropic body-wave field measuring displacement along the x-axis. The CSD is real-valued for isotropic fields.}
\label{fig:css-corr}
\end{figure}

Sufficiently far underground, we have two kinds of seismic waves: compressional waves (also called primary waves or P waves) and shear waves (also called secondary waves or S waves). Having two kinds of body waves reduces the efficiency of the Wiener filter. Because of their different propagation velocity in the ground, P and S waves produce two-point correlations that are out of phase affecting the configuration of the optimal array. The isotropic CSDs are shown as a function of seismometer separation in figure (\ref{fig:css-corr}).

We now consider the example of an isotropic, homogeneous seismic field. Correlation functions (CSDs) between seismometers and with the associated gravity fluctuations can be calculated analytically. The CSD between two seismometers placed at $\vec{r}_1$ and $\vec{r}_2$ due to an isotropic compressional-wave field is given by:
\begin{equation}
\begin{split}
\langle(\vec{e}_1 & \cdot\vec{\xi}^{\,\rm P}(\vec{r}_1,\omega)),(\vec{e}_2\cdot\vec{\xi}^{\,\rm P}(\vec{r}_2,\omega))\rangle \\
&=S(\xi^{\rm P},\omega)\left[(j_0(\Phi_{12})+j_2(\Phi_{12}))(\vec{e}_1\cdot\vec{e}_2) -3j_2(\Phi_{12})(\vec{e}_1\cdot\vec{e}_{12})(\vec{e}_2\cdot\vec{e}_{12})\right],
\end{split}
\end{equation}
with $\Phi_{12}\equiv k^{\rm P}|\vec r_2-\vec r_1|$ ($k^{\rm P}$ being the wave number of P waves), and for an isotropic shear-wave field:
\begin{equation}
\begin{split}
\langle(\vec{e}_1&\cdot\vec{\xi}^{\,\rm S}(\vec{r}_1,\omega)),(\vec{e}_2\cdot\vec{\xi}^{\,\rm S}(\vec{r}_2,\omega))\rangle \\
&=S(\xi^{\rm S},\omega)\left[(j_0(\Phi_{12})-\frac{1}{2}j_2(\Phi_{12})) (\vec{e}_1\cdot\vec{e}_2) +
\frac{3}{2}j_2(\Phi_{12})(\vec{e}_1\cdot\vec{e}_{12})(\vec{e}_2\cdot\vec{e}_{12})\right],
\end{split}
\end{equation}
with $\Phi_{12}\equiv k^{\rm S}|\vec r_2-\vec r_1|$ ($k^{\rm S}$ being the wave number of S waves). In the last two equations, $\vec{e}_1$ and $\vec{e}_2$ are the directions of the measurement axes of the two seismometers, $\vec{e}_{12}$ is the unit separation vector between them, and $S(\xi^{\rm P},\omega)$ and $S(\xi^{\rm S},\omega)$ are the spectral densities of the P and S waves, respectively.

An off-diagonal component of the matrix of seismometer CSDs can then be written:
\begin{equation}
C_{\rm SS}^{12} = \langle(\vec{e}_1\cdot\vec{\xi}^P(\vec{r}_1,\omega)),(\vec{e}_2\cdot\vec{\xi}^P(\vec{r}_2,\omega))\rangle + \langle(\vec{e}_1\cdot\vec{\xi}^S(\vec{r}_1,\omega)),(\vec{e}_2\cdot\vec{\xi}^S(\vec{r}_2,\omega))\rangle
\label{eq:Css}
\end{equation}
Here, we want to introduce the total seismic spectral density $S(\xi,\omega)$ and define $S(\xi^{\rm P},\omega)=pS(\xi,\omega)$ and $S(\xi^{\rm S},\omega)=(1-p)S(\xi,\omega)$, where $p$ is the polarization mixing parameter. It should be noted that the last equation assumes that the CSDs between P and S-wave displacements are negligible.

The diagonal of the CSD matrix $C_{\rm SS}$ contains the spectral densities of all seismometers. One therefore needs to add a contribution to the diagonal coming from the seismometers' instrument noise. This can be achieved by multiplying the seismic spectral density on the diagonal by $(1+1/{\rm SNR}^2)$. Since the seismic field is homogeneous, all seismometers observe the same seismic spectral density, and if all seismometers have the same sensitivity, then the values on the diagonal of $C_{\rm SS}$ are all the same and equal to $S(\xi,\omega)(1+1/{\rm SNR}^2)$.

In order to calculate the CSD between the test-mass acceleration $\delta \vec{a}$ due to gravity fluctuations (NN) and the seismometers, we first need a gravitational coupling model. As mentioned earlier, the test mass is assumed to be located in a underground cavity. For simplicity, the cavity has a spherical shape and the test mass is at its center. Also, it is assumed that the cavity has a small radius $a$ so that $k^{\rm P}a\ll1$ and $k^{\rm S}a\ll1$. The last conditions will clearly by fulfilled in underground detectors with typical body-wave speeds of a few km/s. This greatly simplifies the equations describing the gravitational coupling between seismic field and test mass, and also makes sure that we do not need to consider seismic waves scattered from the cavity \cite{Har2015}. If the test mass is not located at the center of the cavity, then the coupling will obtain an additional negligible phase term. The impact of the shape of the cavity volume on the gravitational coupling between seismic field and test mass has not been investigated yet. Under these conditions, the seismic gravity perturbation reads \cite{Har2015}
\begin{equation}
\delta \vec{a}(\omega) = \frac{4}{3}\pi G\rho_0(2\vec{\xi}^{\,\rm P}(\omega)-\vec{\xi}^{\,\rm S}(\omega))
\end{equation}
The corresponding CSD between a seismometer at $\vec r_1$ and the acceleration of the test-mass located at $\vec r_0$ along direction $\vec e_{\rm tm}$ is:
\begin{equation}
C_{\rm SN} = \frac{4}{3}\pi G\rho_0\big[2\langle (\vec{e}_{\rm tm}\cdot \vec{\xi}^{\,\rm P}(\vec{r}_0,\omega)),(\vec{e}_1\cdot \vec{\xi}^{\,\rm P}(\vec{r}_1,\omega))\rangle - \langle (\vec{e}_{\rm tm}\cdot \vec{\xi}^{\,\rm S}(\vec{r}_0,\omega)),(\vec{e}_1\cdot \vec{\xi}^{\,\rm S}(\vec{r}_1,\omega))\rangle\big]
\label{eq:Csn}
\end{equation}
In a similar way, we can calculate the spectral density of the test-mass acceleration:
\begin{equation}
\begin{split}
C_{\rm NN} &= \left(\frac{4}{3}\pi G\rho_0\right)^2\big[4\langle (\vec{e}_{\rm tm}\cdot \vec{\xi}^{\,\rm P}(\vec{r}_0,\omega))^2\rangle + \langle (\vec{e}_{\rm tm}\cdot \vec{\xi}^{\,\rm S}(\vec{r}_0,\omega))^2\rangle\big]\\
&= \left(\frac{4}{3}\pi G\rho_0\right)^2S(\xi;\omega)(3p+1)
\end{split}
\label{eq:Cnn}
\end{equation}
All terms inside equation (\ref{eq:residual}) are now available as analytic expressions for the isotropic seismic field, which allows us to study cancellation performance of Wiener filters using seismometer arrays and to search for optimal array configurations. 

\section{Optimization of seismic arrays for NN cancellation}
\label{sec:optim}
\subsection{Validation of algorithms using the case of surface Rayleigh waves}
The optimization of seismometer arrays for isotropic Rayleigh-wave fields was addressed in previous publications \cite{DHA2012,CoEA2016a}. We have used this case to validate our optimization algorithms, which however requires a different tuning of certain parameter settings of the optimization method. Consistency with previous results was achieved, and in the following, we show the results of a new analysis of the robustness of the cancellation performance with respect to small deviations of the seismometer locations from their optimum.
\begin{figure}[t]
\centering\includegraphics[width=0.8\textwidth]{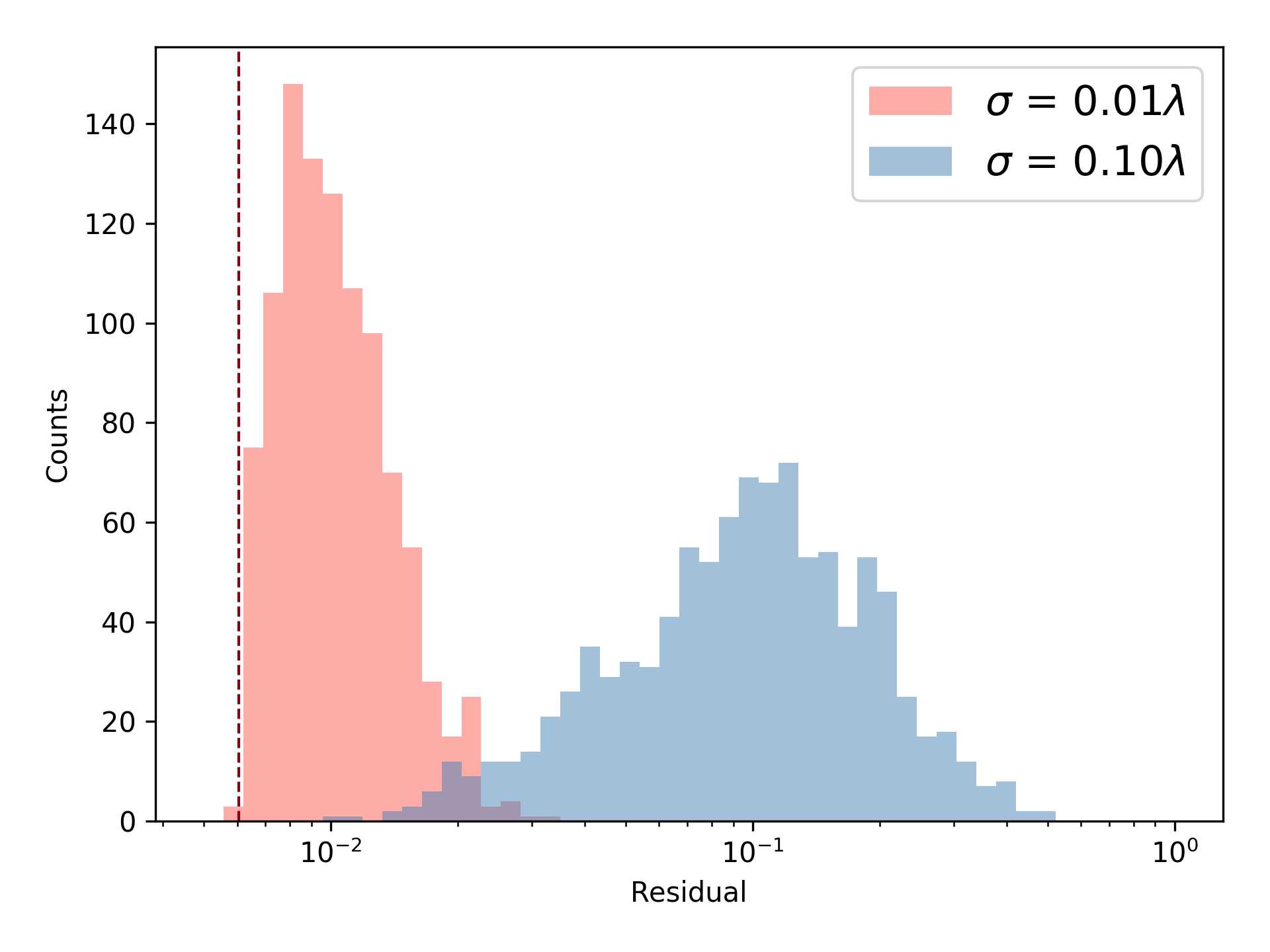}
\caption{Robustness of cancellation performance for NN from Rayleigh waves determined by random Gaussian shifts of seismometer coordinates from their optimal values. The array contains $N=6$ sensors with $\rm SNR=100$.}
\label{fig:rayleigh}
\end{figure}

\begin{figure}[t]
\centering\includegraphics[width=0.8\textwidth]{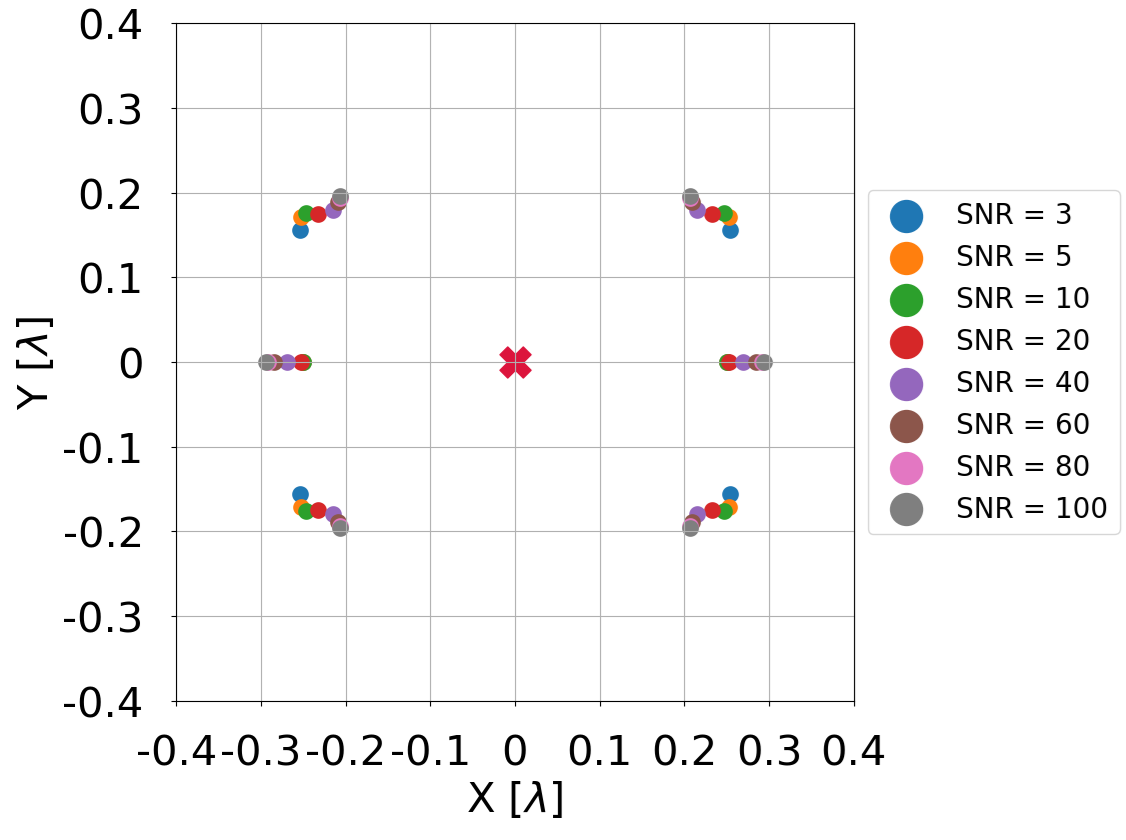}
\caption{Optimal $N=6$ array configurations with varying seismometer SNR.}
\label{fig:raySNR}
\end{figure}

The case presented here is for $N=6$ seismometers with $\rm SNR=100$ located on a flat surface monitoring an isotropic, homogeneous Rayleigh-wave field. The optimal sensor coordinates can be found in section 7.1.6 of \cite{Har2015}. In the optimal configuration, the approximate distance between seismometers and test mass is about $0.3\lambda$, where $\lambda$ is the length of the Rayleigh waves at the optimization frequency. This distance depends on the seismometer SNR (see figure \ref{fig:raySNR}).

We add a random number to each of the two sensor coordinates of all sensors. The random numbers are drawn from zero-mean Gaussian distributions for two standard deviations: $\sigma=0.01\lambda$ and $\sigma=0.1\lambda$. For each of the resulting array configurations, the NN residual is calculated. The values $\sqrt{R}$ are collected in a histogram as shown in figure \ref{fig:rayleigh}. Coordinate mismatches of a bit less than $0.1\lambda$ could be tolerated to achieve a NN reduction by a factor 10 with $\rm SNR=100$ seismometers. At the LIGO and Virgo sites, Rayleigh-wave speeds at 10\,Hz are about 300\,m/s, which means that $0.1\lambda\approx 3\,\rm m$. 

In figure \ref{fig:raySNR}, the optimal arrays for $N=6$ are shown for varying SNR.
The outer four seismometers describe S-shaped trajectories in this plot moving towards smaller ${\rm abs}(x)$-values, and larger ${\rm abs}(y)$-values with increasing SNR. The two seismometers located at $y=0$ move outwards with increasing SNR.

\subsection{Cancellation of NN from body waves}
Finding the optimal array for NN cancellation means to find the configuration that minimizes the residual in equation (\ref{eq:residual}). This kind of calculation becomes very demanding as the number of seismometers increases. We deal with functions in $3N$-dimensional spaces, where $N$ is the number of seismometers. The optimization algorithms do not guarantee to find the global minimum (within a finite time). The optimization of seismometer arrays for body-wave NN cancellation is more demanding than in the case of seismic surface waves for two main reasons: the array needs to disentangle NN contributions from compressional and shear waves, which gives the residual function a richer structure in terms of local minima, and as we will see, there is no unique optimum due to symmetries of the seismic field. 

We used two different global optimizers: Basin Hopping (BH) and Differential Evolution (DE). Basin Hopping is a combination of a local minimizer (gradient descent) with a global Monte Carlo search of the minimum \cite{BH_wales1997,BH_1} based on the Metropolis criterion \cite{Metropolis}, while DE is part of the family of evolutionary algorithms \cite{EvolAlg, DE_Storn1997}. Both algorithms require parameter tuning to efficiently find the global minimum instead of some local minimum with higher noise residuals. For DE, we had to specify the coordinate boundaries to look for the minimum. For BH, we had to specify the temperature and the step size, which serves to explore the space in order to find the minimum.

We adopted three different methods to validate our solutions. First, as mentioned already, we had to achieve a match of the Rayleigh-wave results with the published solutions. Second, we have analytic solutions of the optimal arrays and their residuals for $N=1$ (and arbitrary values of $p$), and for $p=0$ and $p=1$ (for arbitrary number $N$) that can be compared with the numerical solutions. For $p=0$ and $p=1$, the residual is given by equation (\ref{eq:minR}). Third, as explained after equation (\ref{eq:minR}), for sufficiently high $N$, the noise residual $R$ needs to fall at least with $1/N$ when increasing $N$. The results passed all three tests.
\begin{figure}[ht!]
\begin{center}
	\subfloat[Minimum residual $R=0.4298$ found with BH algorithm. Seismometers with one measuring axis along x parallel to the relevant test-mass displacement.]{\includegraphics[width=0.485\textwidth]{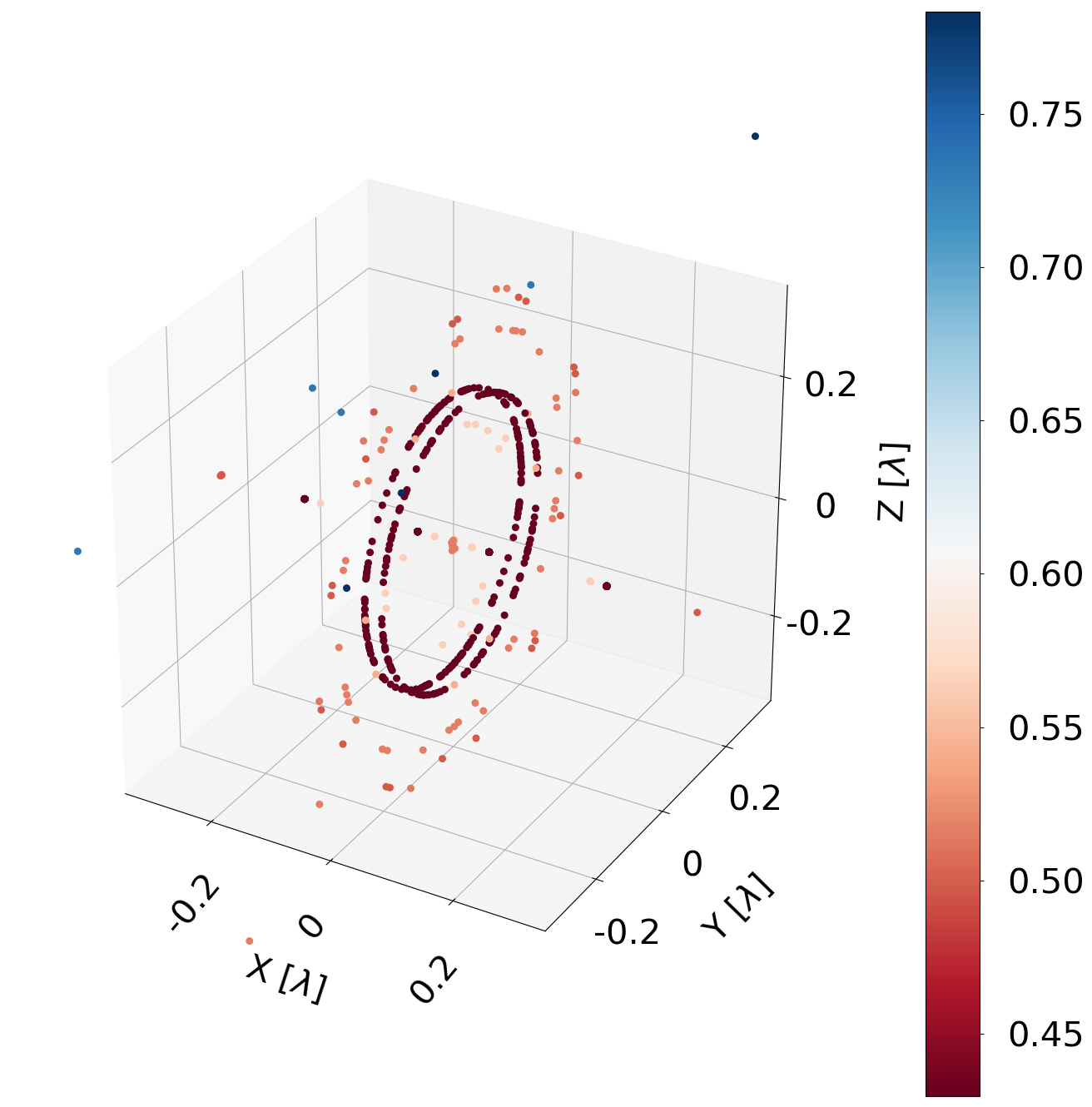}\label{subfig:BHch1}}
	\hfill
	\subfloat[Minimum residual $R=0.4159$ found with BH algorithm. Seismometers with three measuring axes (x,y,z), relevant test-mass displacement along x.]{\includegraphics[width=0.485\textwidth]{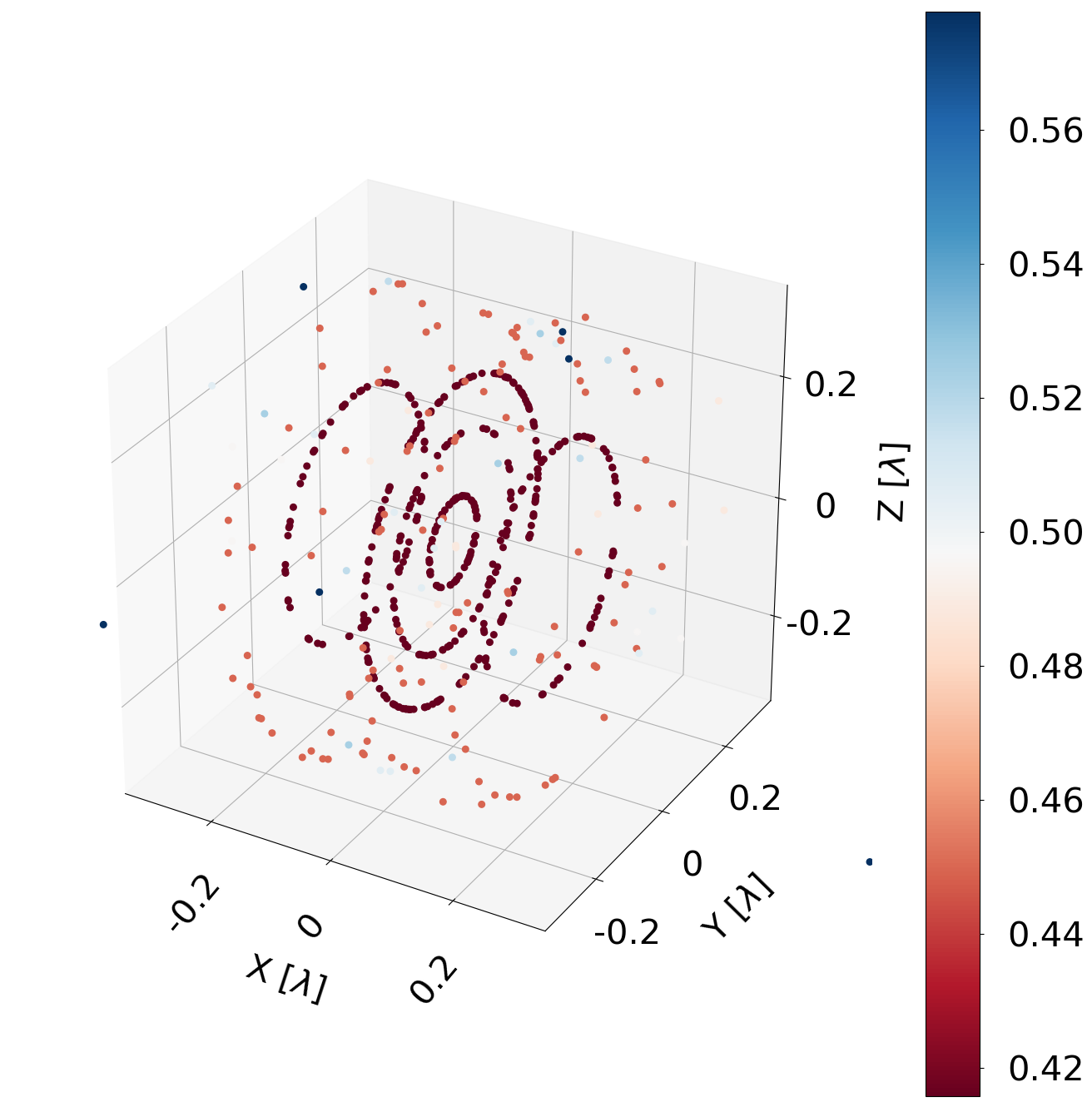}\label{subfig:BHch3}}
	\vskip\baselineskip
	\subfloat[Minimum residual $R=0.4298$ found with DE algorithm. Seismometers with one measuring axis along x parallel to the relevant test-mass displacement.]{\includegraphics[width=0.485\textwidth]{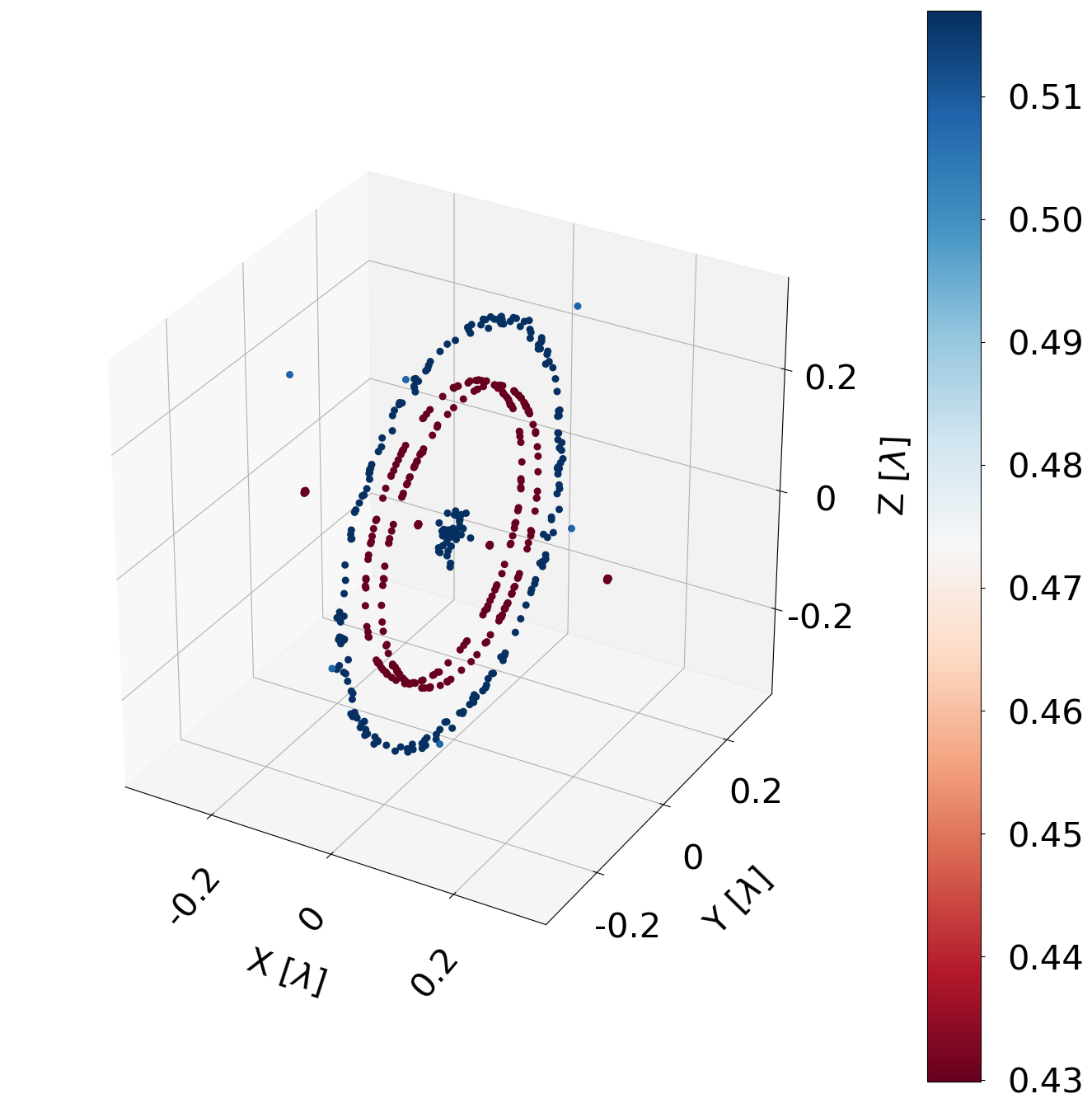}\label{subfig:DEch1}}
	\hfill
	\subfloat[Minimum residual $R=0.4159$ found with DE algorithm. Seismometers with three measuring axes (x,y,z), relevant test-mass displacement along x.]{\includegraphics[width=0.485\textwidth]{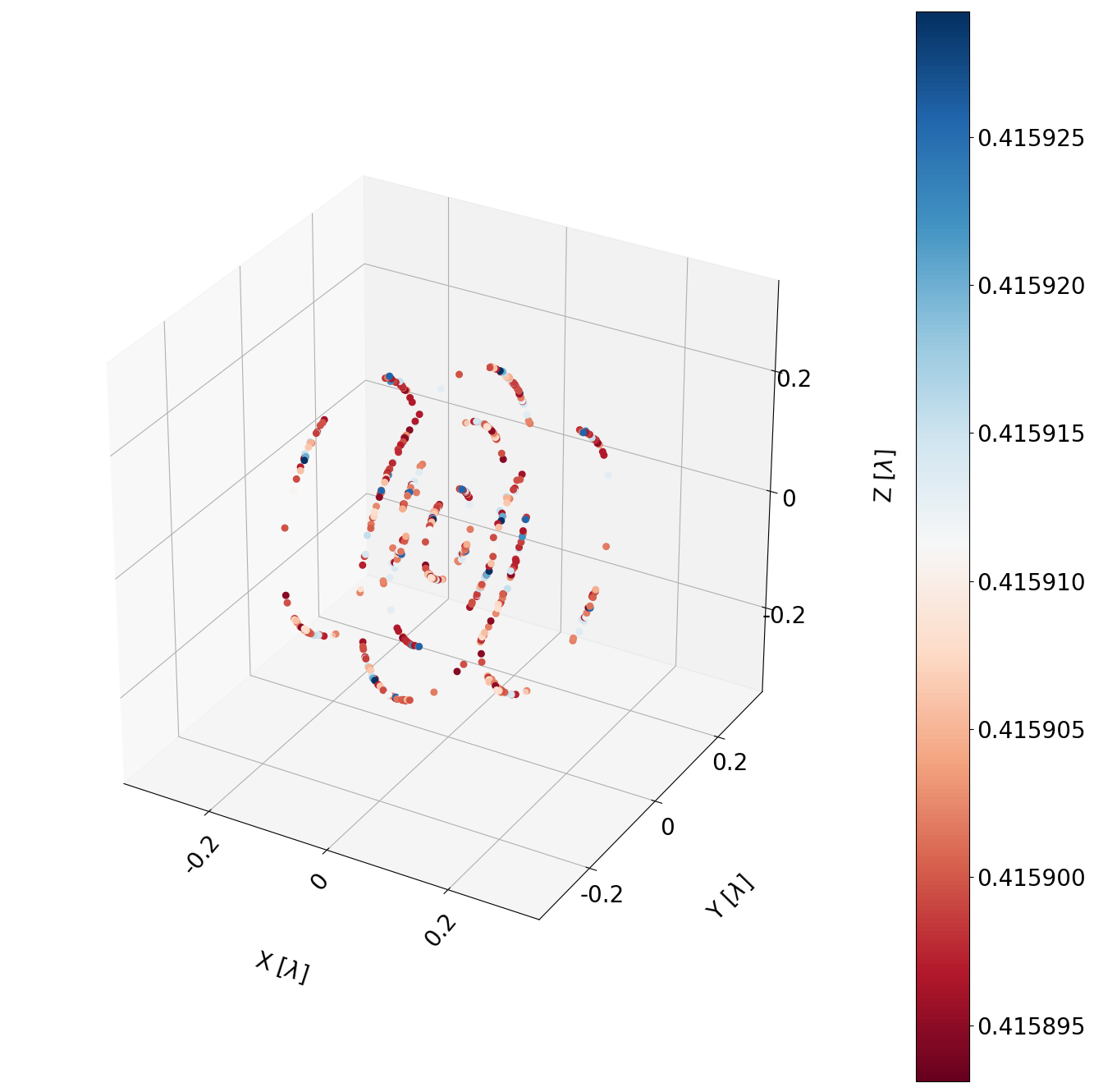}\label{subfig:DEch3}}
	\end{center}
	\caption{Comparison of DE with BH algorithm, and of single-axis with three-axis case. The number of seismometers is $N=6$ with $\rm SNR = 15$. The seismic field has a mixing ratio $p=1/3$.  In each case, the optimization was run 100 times and solutions collected in one plot. The colors on the bar measure the different values of residual obtained with the minimization.} 
\label{fig:DE-BH_tot}	
\end{figure}

Nevertheless, running the global optimizers many times, different solutions are found, many of them corresponding to local minima, others being equivalent optimal solutions due to symmetry. In Figure \ref{fig:DE-BH_tot}, the plots show the seismometer locations for $N=6$ of all solutions from 100 runs. The x-axis corresponds to the direction of the detector arm, which means that it corresponds to the relevant direction of test-mass displacement. We considered two kinds of seismometers: a single-axis seismometer monitoring displacement along the x-axis, and a three-axis seismometer monitoring (x,y,z) displacement. We used $\rm SNR=15$ for all seismometers, and the residual was minimized at a single frequency to give rise to a fixed compressional wavelength $\lambda$ used as length unit in the plots. The corresponding length of shear waves is $\lambda_{\rm s}=0.67\lambda$. The markers are colored according to the residual achieved by the array. 

The DE algorithm performs better on average, but best solutions found with BH and DE over 100 runs perform equally well. Interestingly, the three-axis arrays do not perform significantly better than the single-axis arrays even though the number of channels is $3N$ vs $N$. It means that there is very little extra information that can be extracted from the y,z-axes (the x-axis being the relevant direction of test-mass displacement). 

In table \ref{tab:pos6}, we present values of optimal sensor locations for $N=6$ and $\rm SNR=15$ as a benchmark. These results were obtained with DE using decreased tolerances on the sensor positions to give precise values up to 5 decimal places while larger tolerances are acceptable (and used throughout the rest of the paper) to get arrays with very similar configuration and performing equally well for all practical purposes.
\begin{table}[ht]
\renewcommand{\arraystretch}{1.5}
\centering{
\begin{tabular}{lp{5cm}l}
\hline
Configuration & Sensor coordinates [$\lambda$] & Noise residuals $\sqrt{R}$\\
\hline
Single-axis & (-0.014,\,0,\,-0.224),\newline
			  (-0.014,\,0,\,0.224), \newline
   			  (-0.014,\,-0.224,\,0),\newline
			  (-0.014,\,0.224,\,0),\newline   			  
   			  (0.059,\,0,\,0),\,(0.250,\,0,\,0) & 0.430\\
Three-axis &  (0.152,\,0,\,0.183),\newline
			  (-0.152,\,0,\,0.183),\newline
   			  (0,\,-0.230,\,0.040),\newline
   			  (0,\,0.230,\,0.040),\newline
   			  (0,\,0,\,0.064),\,(0,\,0,\,-0.158) & 0.416\\
   			  
\hline
\end{tabular}}
\caption{Benchmark solution for $N=6$, $\rm SNR=15$, and $p=1/3$. Coordinates are given in units of compressional-wave length $\lambda$.}
\label{tab:pos6}
\end{table}

\begin{figure}[htbp]
	\centering\includegraphics[width=0.80\textwidth]{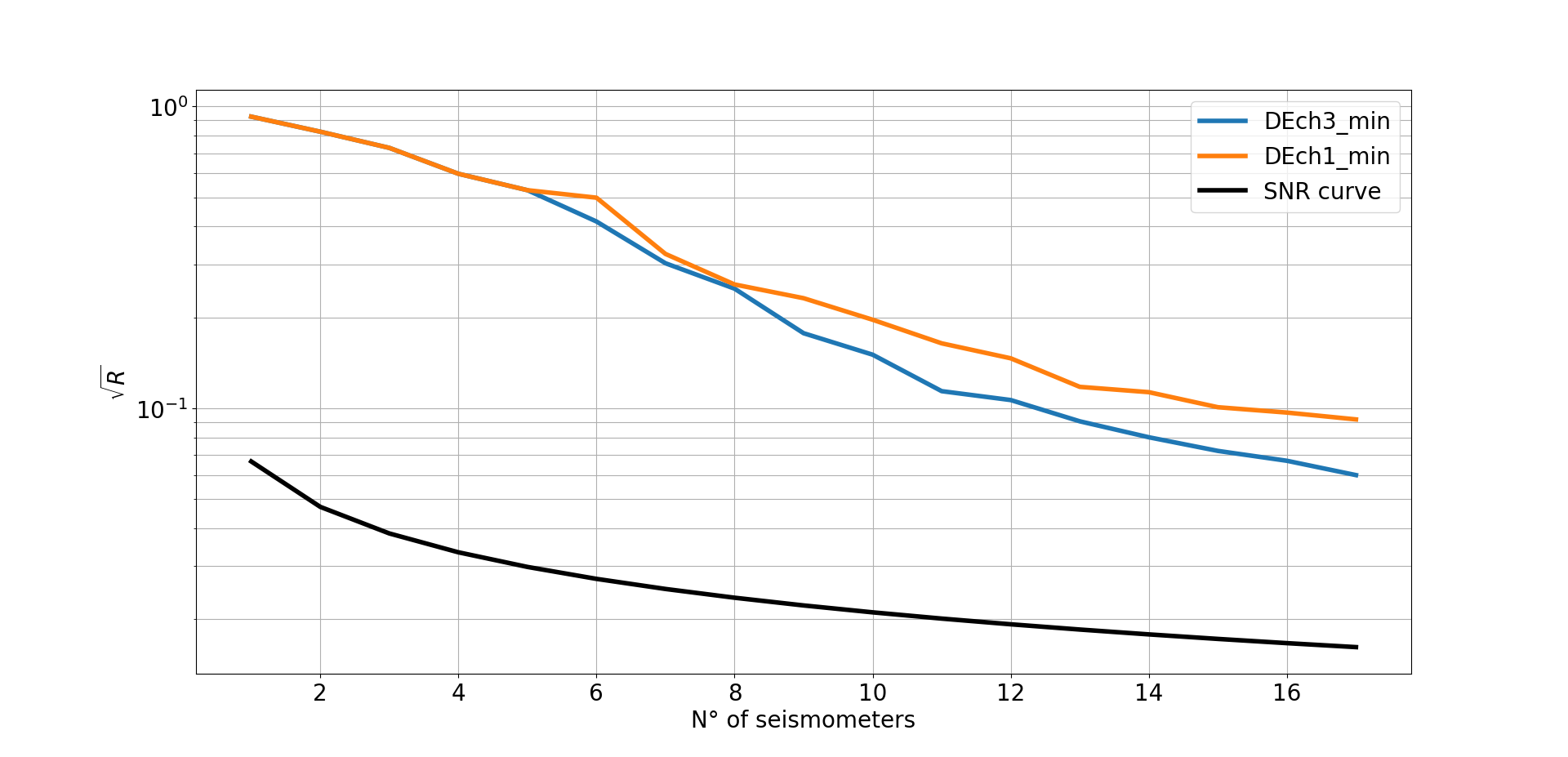}
	\caption{Results obtained with DE with three-axes and single-axis sensors and $p=1/3$. The theoretical sensor-noise limit is shown as black curve (SNR curve). The residuals correspond to the minimum over 100 optimization runs for each number of seismometers.}
	\label{fig:R_vs_N}
\end{figure}
The solutions for the single-axis and three-axis solutions are not unique. Any rotation of the array around the x-axis yields another solution with the same noise residual. Therefore, these benchmark values are obtained by taking the result of the optimization and rotating it such that symmetry axes are aligned with coordinate axes. In figure \ref{fig:R_vs_N}, the residuals are shown for the single and three-axis sensors as a function of $N$. A residual of $\sqrt{R}<0.1$ is achieved for $N>14$. For comparison, the plot also shows the theoretical sensor-noise limit from equation (\ref{eq:minR}). At high $N$, the curves start to fall with similar slope, which means that any new sensor just serves to effectively improve the sensitivity of the array without significantly affecting the Wiener filter's ability to disentangle different modes and polarizations of the field.

We investigate the robustness of cancellation performance with respect to shifts in sensor locations from their optimum. Random numbers drawn from a zero-mean Gaussian are drawn with two standard deviations, $\sigma=0.01\lambda$ and $\sigma=0.07\lambda$. 
\begin{figure}[htbp]
	\centering
	\includegraphics[width=0.8\textwidth]{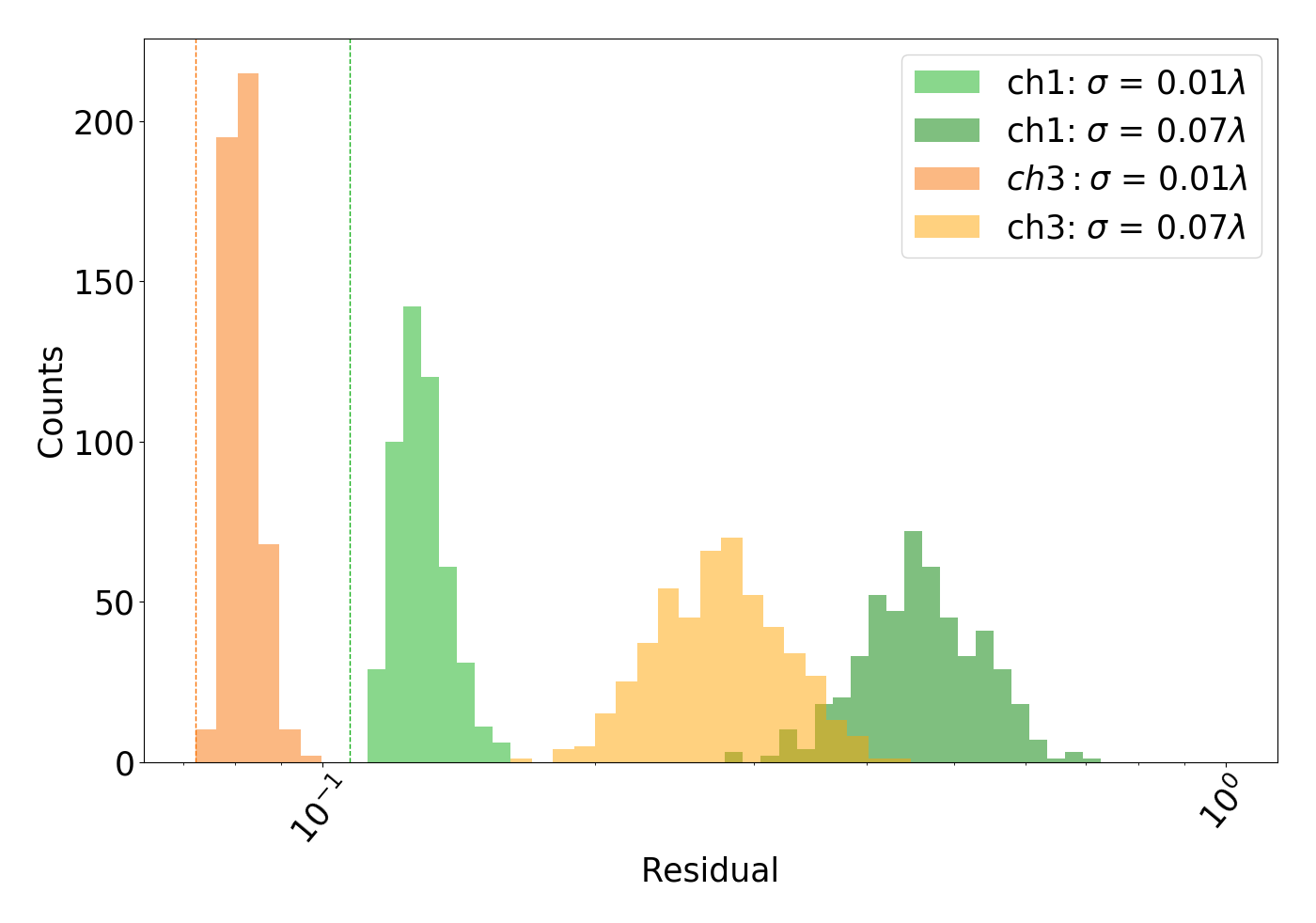}
	\caption{The histograms show the variability of the residual functions with $N=15$ seismometers (single and three-axis) if we modify the optimized array coordinates with a Gaussian with standard deviation of $\sigma$. The two vertical lines show the residual for the optimized coordinates.}
\label{fig:histo}
\end{figure}
As shown in figure \ref{fig:histo}, sensor coordinates for an array of three-axis seismometers can deviate by (in average) $0.07\lambda$ from their optimal values to achieve a factor 3 reduction of body-wave NN. We argue in the next section that a factor 3 NN suppression is likely sufficient to achieve sensitivity targets of the future GW detector Einstein Telescope.

\section{Implications for the Einstein Telescope}
\label{sec:ET}
\begin{figure}[htbp]
\begin{center}
	\subfloat[ET at the surface.]{\includegraphics[width=0.485\textwidth]{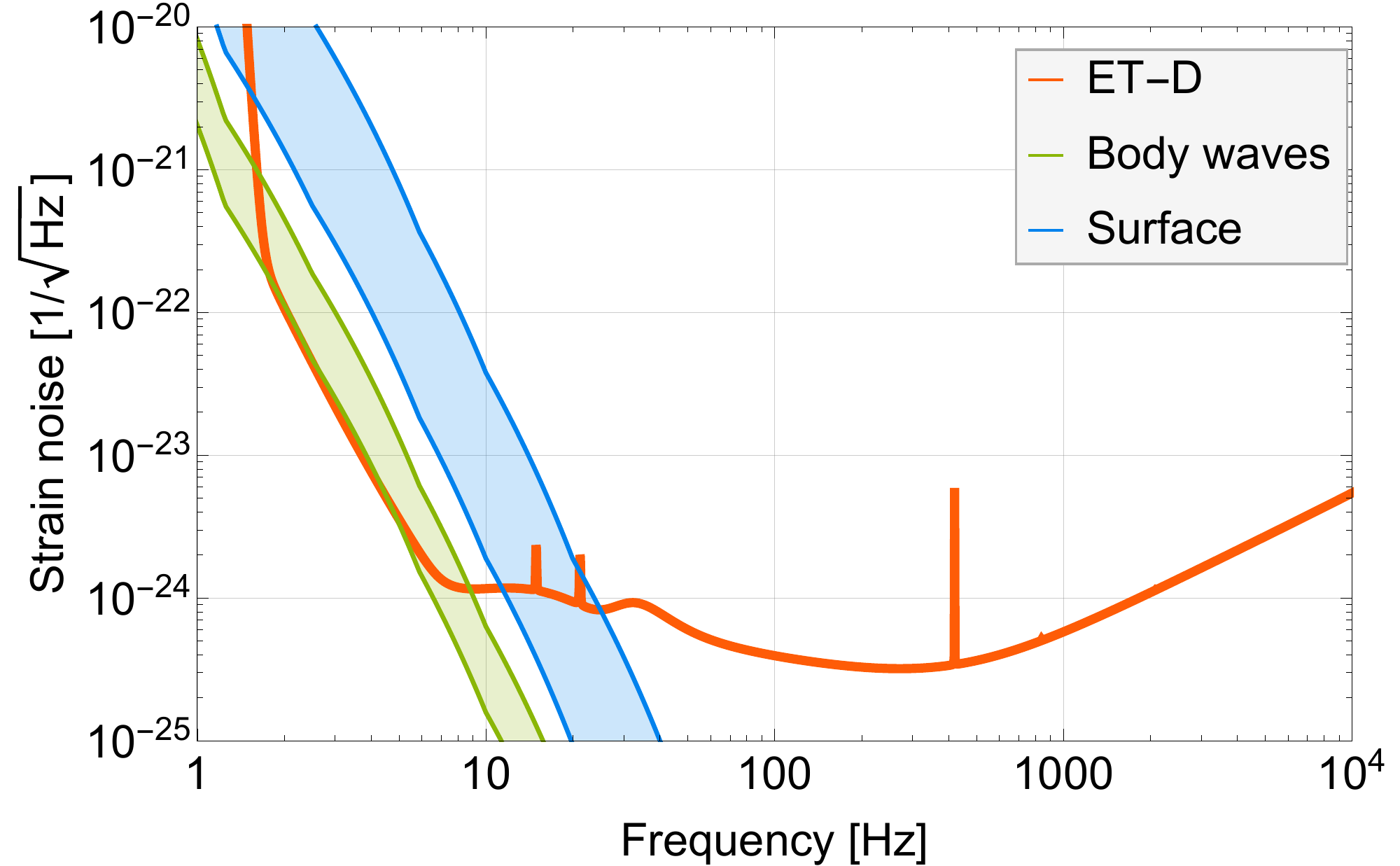}}
	\hfill
	\subfloat[ET underground.]{\includegraphics[width=0.485\textwidth]{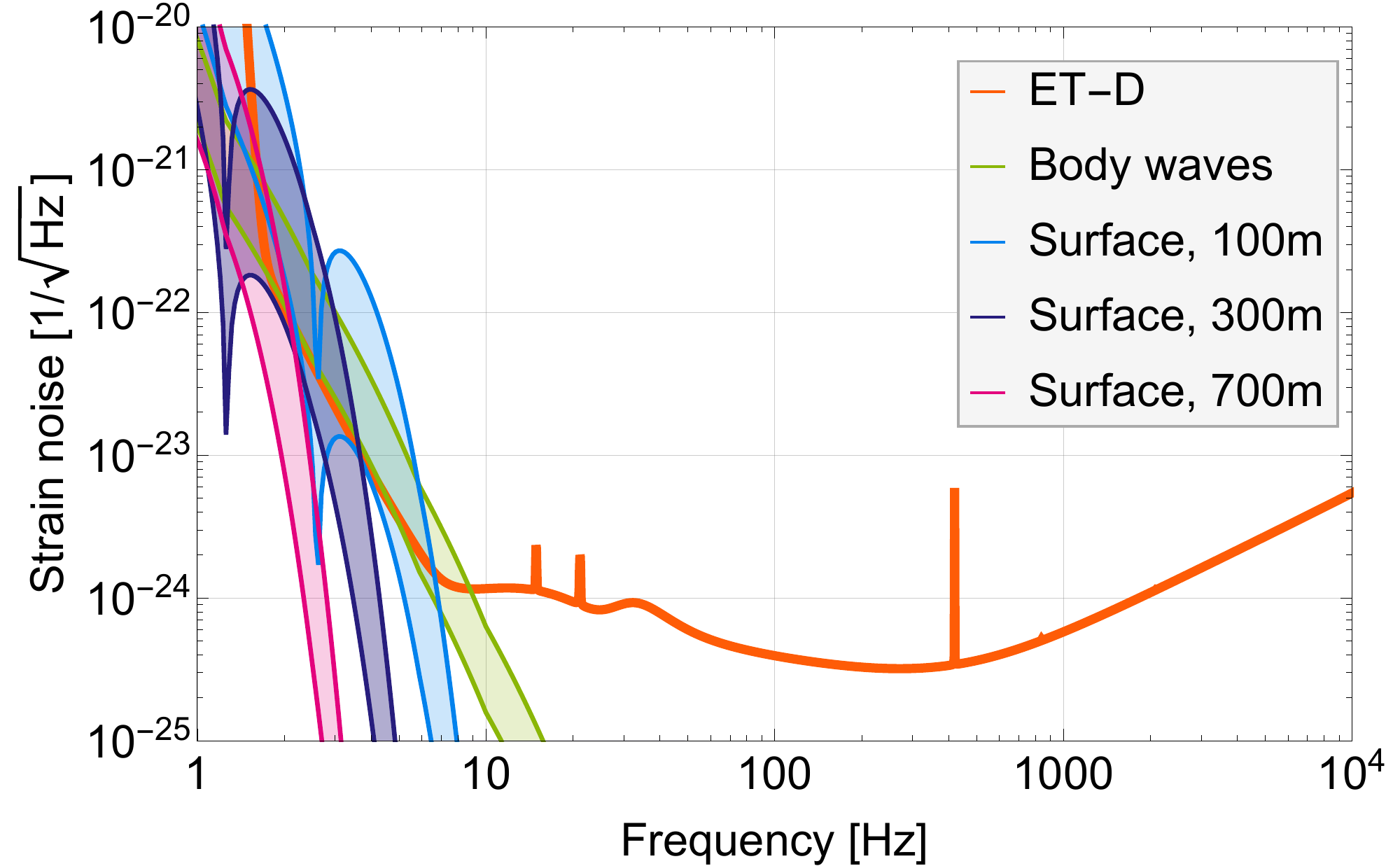}}
	\end{center} 
	\caption{Seismic NN predictions for the Einstein Telescope.}	
	\label{fig:ET}
\end{figure}

The Einstein Telescope (ET) is a proposed third-generation GW detector to be constructed underground \cite{PuEA2010}. The main motivation for underground construction is to strongly suppress NN from atmospheric \cite{FiEA2018} and seismic fields \cite{BeEA2010}. However, NN will still play a role. As shown in figure \ref{fig:ET}, NN from surface waves will be insignificant if the detector is constructed a few 100\,m underground. However, the NN from seismic body waves cannot be avoided at any depth, and it becomes a sensitivity-limiting noise contribution below 10\,Hz. Depending on the quality of the underground site, one still needs to mitigate body-wave NN by up to a factor 10. 

The range of body-wave NN shown in the two plots assumes that underground seismic spectra are a factor 3 to 12 above the global low-noise model \cite{Pet1993}, and an isotropic field is composed entirely of compressional waves. If it were composed entirely of shear waves, then the NN would be a factor 2 smaller. The prediction of Rayleigh NN (denoted {\it Surface} in the two plots) in underground detectors requires an assumption about the seismic surface spectrum, which is a factor 50 to 1000 above the global low-noise models in the two plots, but also an assumption about the dispersion curve. The slower (and therefore shorter) Rayleigh waves, the stronger is the suppression of associated NN with depth \cite{Har2015}. 
The dispersion model used for the two plots (Rayleigh wavelength plays a negligible role for NN in surface detectors) yields a Rayleigh-wave speed of 1.8\,km/s at 1\,Hz falling to 500\,m/s at 10\,Hz. There can be significant regional variations, but these values are typical. The NN model takes into account for body-wave as well as Rayleigh-wave NN that there are additional contributions from cavity-wall displacement (combining Sections 3.3.1 and 3.4.2 in \cite{Har2015}). For the body-wave and Rayleigh-wave field, anisotropy can increase or decrease NN relative to the isotropic level shown in figure \ref{fig:ET}.

It follows that planning for ET must include NN cancellation, and it will be essential to have a detailed understanding of the seismic field in terms of spectra, speeds, or more accurately, two-point spatial correlations. This will make possible the precise prediction of NN in the underground detector, which determines the required NN cancellation, and the calculation of optimal sensor locations. Due to the uncertainties of the models of the seismic fields used in figure \ref{fig:ET}, i.e., lacking a detailed understanding of the seismic field, NN suppression of up to a factor 10 are potentially required, but if the underground site is among the quieter ones, factor 3 is very likely sufficient.

\begin{figure}[htbp]
\begin{center}
	\includegraphics[width=0.7\textwidth]{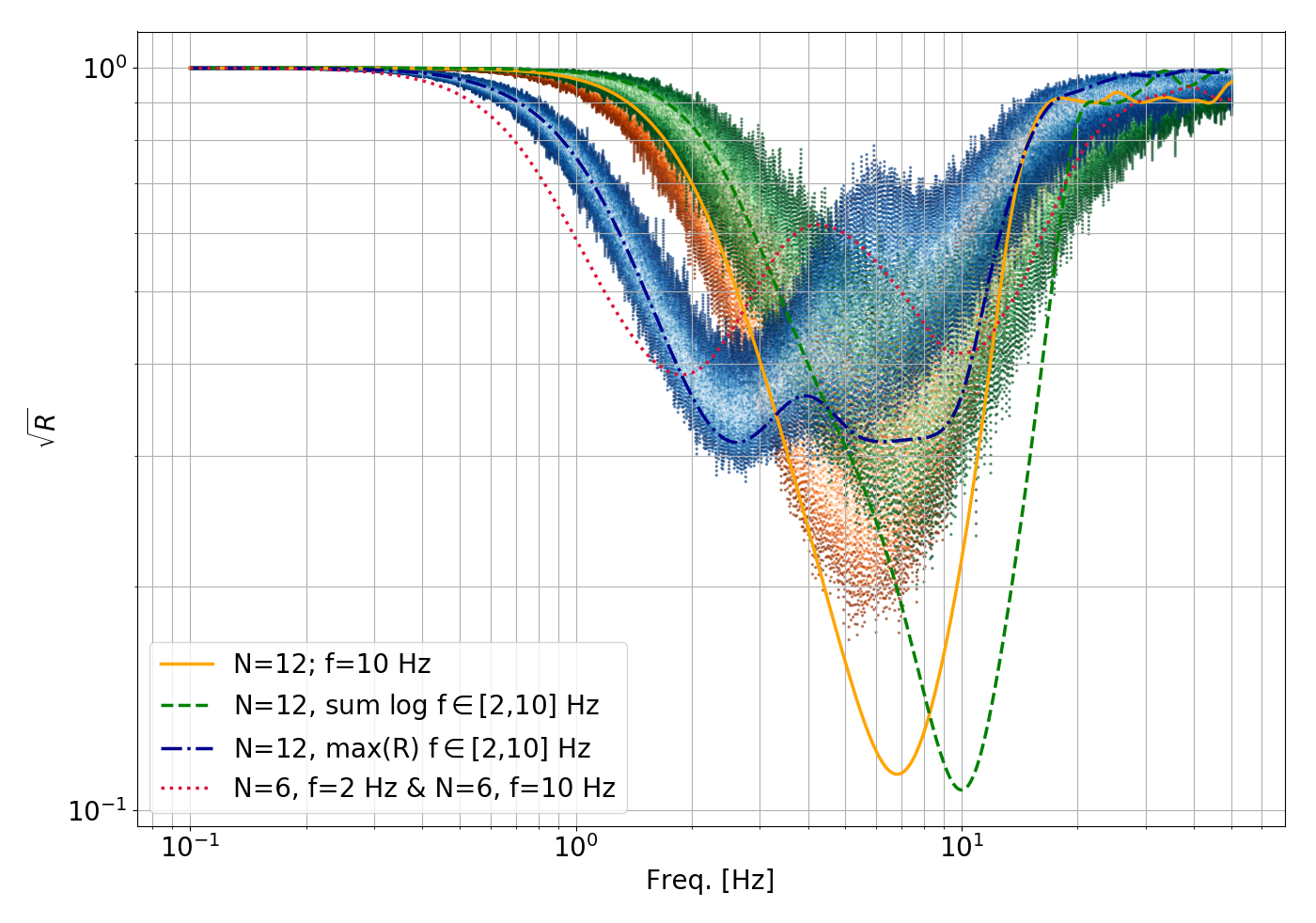}
	\end{center} 
	\caption{Broadband optimization of cancellation performance including histograms of residuals if all sensor coordinates are changed randomly by adding a number drawn from a Gaussian with standard deviation of 50\,m corresponding to about $0.08\lambda$ at 10\,Hz. The curve $N=6+6$ is plotted without histogram.}	
	\label{fig:broadband}
\end{figure}
One additional aspect of noise cancellation is the width of the frequency band over which it is effective. According to the noise plots in figure \ref{fig:ET}, cancellation for ET might be required over a larger band of frequencies between about 2\,Hz and 10\,Hz. Figure \ref{fig:broadband} shows the results of various attempts to achieve broadband-optimized cancellation. The solid curve shows the NN reduction using an array with 12 seismometers optimized at 10\,Hz. In this case, NN reduction at 2\,Hz is minor. Similar performance is obtained when minimizing the sum of residuals between 2\,Hz and 10\,Hz using 12 seismometers as shown by the dashed curve. The dotted curve results from merging two arrays with 6 seismometers each, one optimized for 2\,Hz, the other for 10\,Hz. Good suppression can be achieved at low and high frequencies, but the performance is not uniform over the entire NN band. The best solution was found by minimizing the maximum residual over the band 2\,Hz to 10\,Hz (dot-dashed). It is likely that other cost functions can be found to yield even better results. 

The optimization results presented in this paper are only indicative of course. The seismic field will not be isotropic, homogeneous. Nonetheless, isotropic fields pose a greater challenge to NN cancellation designs than anisotropic fields \cite{CoEA2016a}, and since inhomogeneities are caused by the presence of local sources or strong scattering of seismic waves, it is possible to adapt the array provided that the location of local sources and scattering centers are known. An important result from section \ref{sec:NNcanc} is that seismometer positions do not need to be exactly matching the optimal positions (see figure \ref{fig:histo}). Even strongly degraded configurations with respect to the optimum still achieve a factor 3 reduction of NN in our analysis. We therefore conclude that reduction of NN in ET by a factor 10 using coherent cancellation of body-wave NN would be feasible. Clearly, it remains a significant effort since boreholes for about 15 seismometers per test mass need to be drilled and a site-characterization campaign is required to obtain two-point spatial correlations of the seismic field. It should also be mentioned that the cancellation can be achieved with already existing commercial seismometers, which have instrumental noise below the seismic global low-noise model up to 10\,Hz.

\section{Conclusion}
\label{sec:concl}
We have analyzed the performance of optimized seismometer arrays for the cancellation of body-wave NN using Wiener filters. We found that about 15 sensors are required to reduce NN by a factor 10 (in amplitude) when 1/3 of the spectral density of the seismic field is in compressional waves (the rest being in shear waves). The optimal array configurations were determined for isotropic, homogeneous fields. The cancellation performance is mainly limited by the array's ability to disentangle shear from compressional waves. In contrast, cancellation performance is limited by the seismometer noise if only one wave polarization (either compressional or shear) is present in the seismic field. 

We then found that for a well performing array, the seismometer locations do not have to match the optimal locations precisely. This is true for Rayleigh-wave and body-wave NN cancellation given the respective NN suppression targets in future detectors. 

Cancellation of NN will likely be required to achieve ET sensitivity (according to the reference sensitivity ET-D). Our results lead us to the conclusion that NN cancellation is feasible for underground detectors. Neither the required number of seismometers, nor their sensitivity, nor the required accuracy of their positioning in boreholes is prohibitive. We therefore propose coherent cancellation of NN using Wiener filters as technique in the third-generation GW detector Einstein Telescope. 

\ack{We thank the gravitational-wave group at the Gran Sasso Science Institute for helpful discussions during our meetings. We are also grateful to Luca Rei, the administrators of the INFN/CNAF computer cluster at Bologna, and the DIRAC developers whose help was essential to prepare the computing environment for our optimization algorithms and to make effective use of computing resources.}

\appendix
\section{Optimal arrays for 6 seismometers}
\begin{figure}[ht!]
\begin{center}
	\subfloat[Single-axis sensors.]{\includegraphics[width=0.485\textwidth]{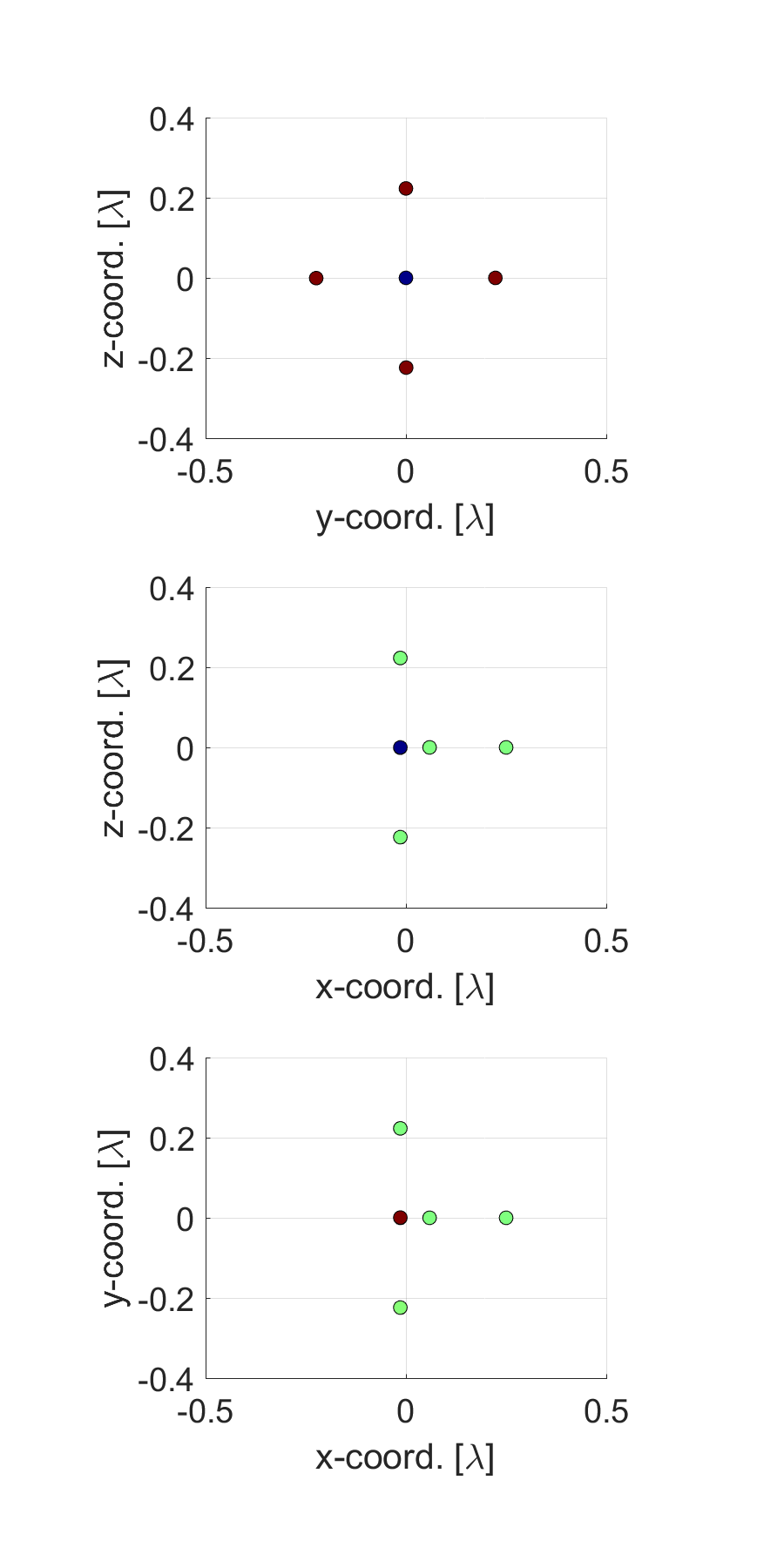}}
	\hfill
	\subfloat[Three-axis sensors.]{\includegraphics[width=0.485\textwidth]{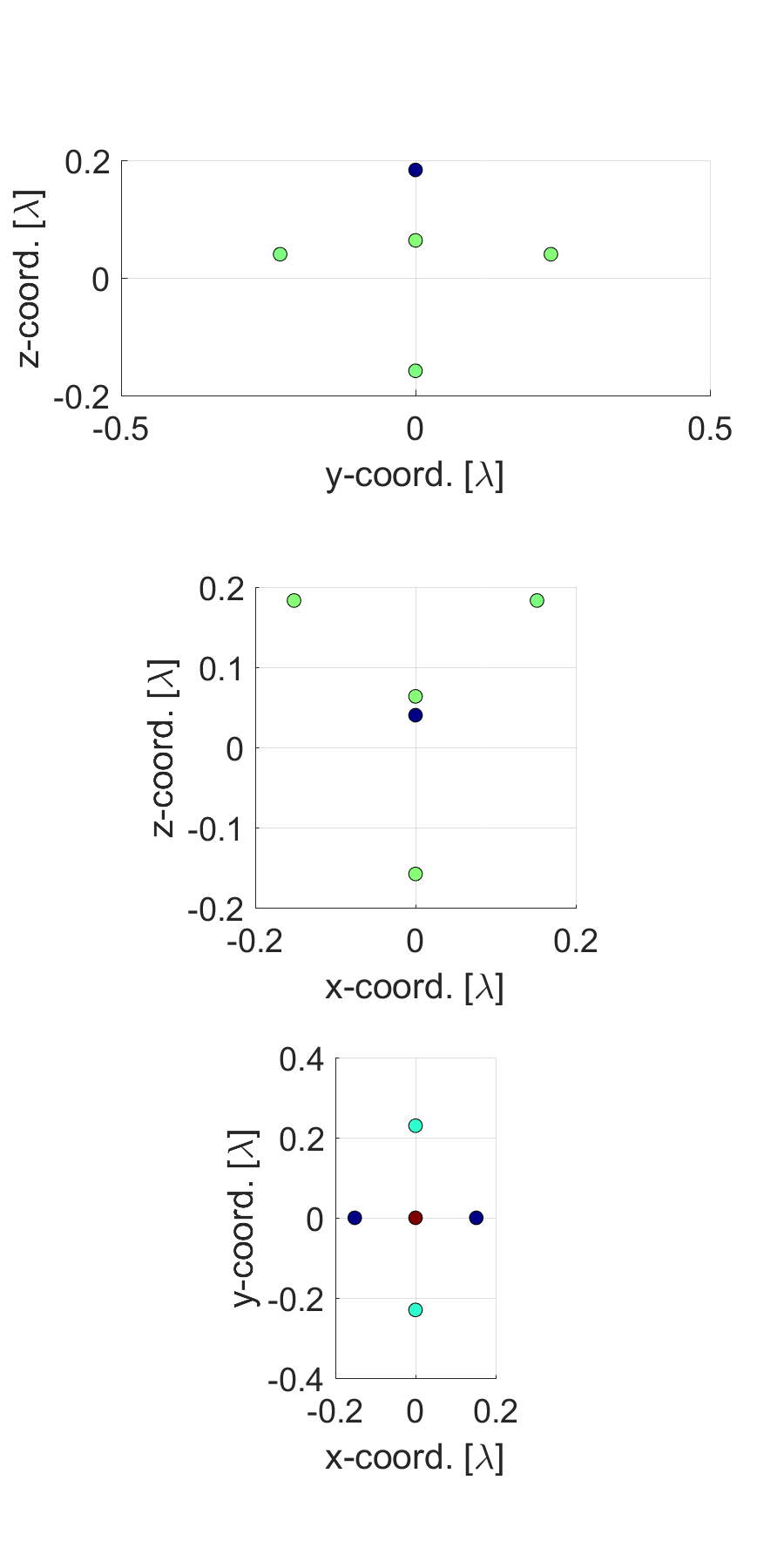}}
	\end{center}
	\caption{Optimal array configurations for $N=6$ and $p=1/3$ (coordinates listed in table \ref{tab:pos6}). Difference in marker colors indicates different values of the coordinate along the projection direction.} 
\label{fig:optarray}	
\end{figure}

\section*{References}
\bibliography{Optimize3D}
\bibliographystyle{iopart-num}

\end{document}